\documentclass[epj,nopacs]{svjour}
\usepackage{graphicx}
\usepackage{amsmath,amssymb,color}
\usepackage{bm,braket}
\begin{document}
\title{Novel approach to the removal of the Pauli-forbidden states 
in the orthogonality condition model: A case of multi-$\alpha$ systems}
\author{
  H. Moriya\inst{1}\and 
  W. Horiuchi\inst{2,3,4,1} \thanks{\emph{e-mail:} whoriuchi@omu.ac.jp} \and 
  B. Zhou\inst{5,6}
}
\institute{
Department of Physics, Hokkaido University, Sapporo 060-0810, Japan \and 
Department of Physics, Osaka Metropolitan University, Osaka, 558-8585, Japan \and
Nambu Yoichiro Institute of Theoretical and Experimental Physics (NITEP), Osaka Metropolitan University, Osaka 558-8585, Japan \and
RIKEN Nishina Center, Wako 351-0198, Japan \and 
Key Laboratory of Nuclear Physics and Ion-beam Application (MOE), Institute of Modern Physics, Fudan University, Shanghai 200433, China\and
Shanghai Research Center for Theoretical Nuclear Physics, NSFC and Fudan University, Shanghai 200438, China
}
\date{Received: date / Revised version: date}
\abstract{
We propose to use a basis function
constructed based on the microscopic cluster model for an efficient description
of multi-cluster systems with the orthogonality condition originating from the Pauli principle.
The basis function is expressed analytically by a superposition of correlated Gaussian functions.
We demonstrate the power of this approach by taking an example of a $3\alpha$ system, $^{12}$C.
A comparison with the conventional pseudopotential method using the projection operator is made.
The present method offers efficient and numerically stable computations
as the number of basis functions is significantly reduced compared to the conventional method.
We show that the present basis function includes reasonably small components of the Pauli-forbidden states,
allowing us to discuss simply the structure of the first excited $0^+$ state, Hoyle state.
}
\maketitle

\section{Introduction}

A nuclear clustering phenomenon is an essential aspect of
nuclear structure. In particular, an $\alpha$ ($^4$He) cluster structure
is important to understand the low-lying states of light nuclei~\cite{Ikeda80}.
A striking example includes the first excited $0^+$ state
of $^{12}$C. The so-called Hoyle state~\cite{Hoyle54}
is understood as having a well-developed 3$\alpha$-cluster state
and plays an essential role in nucleosynthesis
to create $^{12}$C element~\cite{Salpeter52}. 
The Hoyle state has attracted interest not
only for astrophysical nuclear physics~\cite{Ogata09,Nguyen12,Ishikawa13,Suno15,Akahori15,Suno16}
but also for the structure itself.
An interpretation of a bosonic condensate state was raised
based on the $\alpha$-cluster model approaches~\cite{Tohsaki01,Funaki03}.
Extending more $\alpha$ cluster systems is highly demanding
in searching for candidates of the condensed states
in 4$\alpha$, 5$\alpha$, and more-$\alpha$ systems
~\cite{Tohsaki01,Yamada04,Wakasa07,Funaki08,Funaki10,Yamada12,Zhou13,Adachi21}.

The structure of a well-clustered system such as the Hoyle state
has been well understood by the microscopic cluster model
calculations~\cite{Uegaki77,Uegaki78,Uegaki79,Kamimura81}.
The microscopic model has an advantage that
the antisymmetrization among nucleons
is fully taken into account. However, an effective nucleon-nucleon
interaction should be taken appropriately for each state and
this adjustment is often complicated.
For example, in the $3\alpha$ and $4\alpha$ cluster models,
the ground state energies of $^{12}$C and $^{16}$O
cannot be reproduced simultaneously
within the nucleon-nucleon interaction level,
implying the need for the breaking effect of $\alpha$ clusters and
an effective three-nucleon interaction~\cite{Itagaki16}.

On the contrary,
the orthogonality condition model (OCM)~\cite{Saito68,Saito69,Saito77}
has been used for studying multi-cluster systems.
In the OCM, clusters are often treated as structureless point particles
and the Pauli principle among the clusters is approximated by
imposing  the orthogonality condition on the redundant bound states
originating from the Pauli principle between clusters,
the Pauli-forbidden states (PFS).
As the OCM starts from the intercluster interaction
that reproduces properties of the subsystems,
it is straightforward to find a reasonable intercluster potential
even if it is phenomenological. Once the Hamiltonian is set, 
the multi-cluster relative wave function
has often been solved with the orthogonality condition
using non-orthogonalized basis expansion
such as Gaussians which generally includes components of the PFS.
To exclude these redundant components from the relative wave function of
the multi-cluster system,
the pseudopotential method~\cite{Kukulin78} has often been used for
practical computations. The pseudopotential comprising
the projection operator to the PFS is added
to the Hamiltonian as a `penalty' function.
By taking its prefactor large,
the PFS are variationally
excluded from the wave function.
Though the implementation of this method is simple,
numerical computations, however, become unstable due to
the large prefactor, typically larger than $10^{3-4}$ MeV.

It is challenging to get a precise solution of
a few-body Schr\"odinger equation with the orthogonality condition.
This sort of problem appears in various cases when a system involves
composite particles~\cite{Varga95,SVMbook,Mitroy13}.
However, it is complicated and a huge number of basis states are required
to get a converged result, see, e.g., Ref.~\cite{Horiuchi14}.
For multi-$\alpha$ systems
the $3\alpha$ system is manageable
with the standard pseudopotential method~\cite{Suzuki08},
although there is some numerical instability in the calculation.
For the $4\alpha$ system, the energy convergence
is extremely slow and a huge number of basis functions
is required~\cite{Funaki08,Orabi11}.
The 5$\alpha$ OCM calculation has not been carried out yet.
Recently, the forbidden-state-free single-particle
basis function was proposed with applications
to the neutron-rich Ca isotopes~\cite{Horiuchi22},
showing great advantages compared to the pseudopotential method
in the stabilization of the numerical computations.
In that case, however, the forbidden states appear only for the
core-nucleon subsystem. Construction of a forbidden-state-free basis
for multi-cluster systems is in general difficult
and challenging problem~\cite{Matsumura06}.

To overcome the difficulties in the removal of the PFS in the OCM approach,
in this paper, we propose an efficient method
for multi-cluster systems using the correlated Gaussian (CG)
expansion~\cite{Varga95,SVMbook}.
The CG expansion method is powerful and flexible to get
a precise solution of the few-body Schr\"odinger
equation not only for nuclear physics but also
other atomic and subatomic fields~\cite{SVMbook,Mitroy13}.
The CG was applied to various multi-cluster systems
and obtained successful results accounting
for light cluster systems~\cite{Suzuki03}.
The authors of Ref.~\cite{Matsumura04}
gave an analytical transform
of the CG to the relative wave function of the microscopic cluster model.
We make use of that transformed CG (TCG)
as a basis for the multi-cluster systems.

In this paper, we take an example of a $3\alpha$ system, $^{12}$C,
where the pseudopotential method can still be handled.
We find that the TCG basis includes the components
of the PFS reasonably small;
no need to introduce the pseudopotential to obtain the physical solutions
of the few-body Schr\"odinger equation.
We discuss the efficiency of the present method for
future applications to more-$\alpha$ systems.

In the next section, theoretical formalism
of the present study is introduced.
In Sec.~\ref{TCG.sec},
we give definitions of a microscopic cluster model
and a basic idea of constructing the present basis function.
The CG is defined and its transformation is recaptured
based on the earlier study~\cite{Matsumura04}.
Section~\ref{three-alpha.sec} defines
the Hamiltonian of the present $3\alpha$ system.
Detailed setups of the present and the ordinary
pseudopotential~\cite{Kukulin78} methods are given.
Section~\ref{sec:resutls} demonstrates
the efficiency of the present basis function 
by comparing it with the conventional pseudopotential method.
The structure of the first excited $0^+$ state is discussed by using
the present basis function. The conclusion
is made in Sec.~\ref{sec:summary}.

\section{Formalism}
\label{sec:formalism}

\subsection{Basis function for the orthogonality condition model}
\label{TCG.sec}

For stable computations of many-body Schr\"odinger equation
with the orthogonality condition,
here we introduce a convenient basis function based on
the microscopic multi-cluster wave function.
We make use of the relative wave function
of the multi-cluster system, i.e.,
the spectroscopic amplitude (SA) of the constituent clusters
as a basis function for the OCM calculation.

\subsubsection{Microscopic multi-cluster model wave function}

Let us start with the definition of the microscopic framework.
A microscopic wave function of 
$N$-cluster ($C_{1}, \dots, C_{N})$ system,
is defined as the antisymmetrized ($\mathcal{A}$) product
of internal cluster wave functions, $\phi(C_i)$,
and the relative wave function among these clusters
$\chi (\bm{x}_1,\ldots \bm{x}_{N-1})$:
\begin{equation}
  \Psi_{\mathrm{micro}} = 
  \mathcal{A} \left[\left(\prod_{i=1}^N\phi(C_i)\right)\chi(\bm{x}_1,\ldots \bm{x}_{N-1})
  \right].
\end{equation}
Indicating $\bm{R}_i$ the position vector of the 
center-of-mass of the $i$th cluster with the mass $m_{C_i}$,
a set of Jacobi coordinates 
is defined by 
\begin{align}
  \begin{pmatrix}
    \bm{x}_1\\
    \bm{x}_2\\
    \vdots\\
    \bm{x}_N\\
   \end{pmatrix}
  =
  \begin{pmatrix}
 1& -1 & 0 &\dots&0\\
 \frac{m_{C_1}}{m_{C_{12}}}&\frac{m_{C_2}}{m_{C_{12}}} &-1& \dots&0\\
 \vdots& \vdots&\vdots &\ddots&\vdots\\ 
 \frac{m_{C_1}}{m_{C_{1\cdots N}}}& \frac{m_{C_2}}{m_{C_{1\cdots N}}}&\dots
 &\dots&\frac{m_{C_N}}{m_{C_{1\cdots N}}}\\
  \end{pmatrix}
  \begin{pmatrix}
    \bm{R}_1\\
    \bm{R}_2\\
    \vdots\\
    \bm{R}_N\\
   \end{pmatrix},
\end{align}
where $m_{C_{1\cdots N}}=\sum_{i=1}^{N}m_{C_i}$.

Following Ref.~\cite{Matsumura04},
the SA of the multi-cluster system is obtained by projecting
the relative wave function
onto a set of coordinates $\bm{t}\equiv (\bm{t}_1,\dots,\bm{t}_{N-1})$
as
\begin{align}
  g(\bm{t})&=\left.\left<\Psi_{\bm t}
  \right|\Psi_{\rm micro}\right>
  \\
  &=\int d\bm{t}'
  \bar{N}(\bm{t},\bm{t}')\chi(\bm{t}') \equiv \mathcal{N} \chi(\bm{t}),
\end{align}
with
\begin{equation}
  \Psi_{\bm t} = \mathcal{A}\left[\left(\prod_{i=1}^N\phi(C_i)\right)
  \left(\prod_{i=1}^{N-1}\delta(\bm{x}_i-\bm{t}_i)\right)\right]
\end{equation} 
where $\left<g|g\right>=\braket{\mathcal{N}\chi|\mathcal{N}\chi}$,
and $\bar{N}(\bm{t},\bm{t}')=\braket{\Psi_{\bm{t}}|\Psi_{\bm{t}'}}$
are respectively called the spectroscopic factor and the norm kernel,
which can be computed systematically, see, e.g., Refs.~\cite{Suzuki03,Tohsaki11}.
Note that the orthogonality condition
\begin{equation}
  \braket{\Psi_{\mathrm{micro}}|\Psi_{\mathrm{micro}}'} 
  =\braket{\chi|\mathcal{N}\chi'}
=
  \begin{cases}
    1  \quad ({\rm for}\ \chi^\prime=\chi)\\
    0  \quad ({\rm for}\ \chi^\prime \neq \chi)\\
  \end{cases}
\end{equation}
is satisfied and hence the normalized
relative wave function
in the microscopic framework should be 
$\mathcal{N}^{1/2}\chi$ but its evaluation
is in general involved.
Here we utilize the SA as a basis function with
$\Psi = g / \sqrt{\braket{g|g}}$. Note that
the orthogonality condition
\begin{align}
  \braket{\Psi|\Psi'}
=
  \begin{cases}
    1  \quad ({\rm for}\ \Psi^\prime=\Psi)\\
    0  \quad ({\rm for}\ \Psi^\prime \neq \Psi)\\
  \end{cases}
  \label{OC2.eq}
\end{align}
is imposed.
In general, Eq.~(\ref{OC2.eq}) is not satisfied in
the microscopic wave functions
but can be a good approximation for a well-clustered state.
In fact, the spectroscopic factor
of $^{12}$C is large about 0.7 and 0.9 for the $0_1^+$ and
$0_2^+$ states, respectively~\cite{Matsumura04}.

\subsubsection{Transformed correlated Gaussian:
  Case of multi-$\alpha$ cluster systems}

In the present study,
we focus on  multi-$\alpha$ systems
although applications to arbitrary cluster systems are straightforward.
Here, each cluster is assumed to have a $(0S)^4$ harmonic oscillator
wave function with the oscillator parameter $\nu$.
In this case, as detailed in Ref.~\cite{Matsumura04},
the SA can be analytically derived by using the integral transform
when $\chi$ is expressed by the correlated Gaussian (CG)~\cite{Varga95,SVMbook}:
\begin{equation}
  G(\nu,A,\bm{x}) = \exp \left(-\frac{1}{2} \nu \tilde{\bm{x}}A \bm{x}\right),
\label{CG.eq}
\end{equation}
where $A$ is a positive definite $(N-1)\times(N-1)$ symmetric matrix.
We define a set of Jacobi coordinates 
excluding the center-of-mass coordinate $\bm{x}_N$ as
$\tilde{\bm{x}}=(\bm{x}_1,\ldots\bm{x}_{N-1})$,
and $\tilde{\bm{x}}A \bm{x}$ is 
a shorthand notation of $\sum_{i,j=1}^{N-1}A_{ij}\bm{x}_{i}\cdot\bm{x}_{j}$,
where a tilde stands for the transpose of a matrix.
The diagonal matrix elements of $A$ can be related to
the geometry of $\alpha$ particles by defining $\nu A_{ii}=1/b_{i}^2$ with
$b_{i}$ being the Gaussian falloff parameter in unit of length,
while $A_{ij}$ describes
the correlation between the $i$th and $j$th particles.
Note that the basis function of Eq.~(\ref{CG.eq})
  is limited to the spin-parity $L^\pi=0^+$ case.
  The extension to more general $L^\pi$ states is
  straightforward by using the global vector
  representation~\cite{Varga95,SVMbook,Suzuki08}.
However, we employ it in the present paper for simplicity.    

The transformed CG (TCG) is again expressed
by a superposition of the CG functions as~\cite{Matsumura04}
\begin{align}
  \bar{G}(A,\bm{x})=& \sum_{p} F_{p}
  \left(\frac{\mathrm{det}(Q+\Gamma)}{\mathrm{det}M_{p}} \right)^{\frac{3}{2}}
  G(\nu,\Gamma -4W_{p}M_{p}^{-1}\widetilde{W_{p}},\bm{x})
\label{TCG.eq}
\end{align}
where $\Gamma$, $Q$, $M_p$, and $W_p$ are
$(N-1)\times (N-1)$ matrices.
$\Gamma$ is the reduced mass matrix
$\Gamma_{ij}=2\mu_i \delta_{ij}$ with 
\begin{equation}
   \mu_{i} = \frac{4i}{i+1}, \qquad (i=1,\cdots,N-1).
\end{equation}
$Q$ and $M_p$ are $(N-1)\times(N-1)$ matrices defined by 
\begin{align}
  Q &= \Gamma (\Gamma-A)^{-1} \Gamma - \Gamma,   \\
  M_{p} &= Q+\frac{1}{2}\Gamma + 2 \widetilde{W_{p}}\Gamma^{-1}W_{p}.
\end{align}
 The coefficient $F_p$ and matrix $W_p$
 can be computed after several algebraic manipulations
 from analytical norm kernels.
 The expression of Eq.~(\ref{TCG.eq}) is nothing but
   the SA of the CG function in which
   the Pauli-principle in the multi-cluster system is incorporated.
See details in Ref.~\cite{Matsumura04}.
Note that $\bar{G}(A,\bm{x})$ is not defined when ${\rm det}(\Gamma-A)=0$
because the integral transform cannot be defined.
For example, this condition holds
when $A_{11}=4$ for a $2\alpha$ system, which corresponds
to the forbidden $0S$ state.
We employ this TCG function $\bar{G}(A,\bm{x})$ as a basis function
for the multi-$\alpha$ system, which takes into account
  the Pauli principle approximately.
The computational cost crucially depends on how many terms
are involved in the summation on $p$ in Eq.~(\ref{TCG.eq}).
The numbers of the terms are 
5, 120, 10147, and 2224955 for $2\alpha$, $3\alpha$, $4\alpha$,
and $5\alpha$ cases, respectively.
The numbers grow rapidly as
the number of $\alpha$ clusters increases
and might be handled up to the $4\alpha$ system.

\subsection{$3\alpha$ cluster model}
\label{three-alpha.sec}

To demonstrate the efficiency of the present basis function,
we take a case of a $3\alpha$ system,
which can be solved by the pseudopotential method~\cite{Kukulin78}.
The Hamiltonian of the $3\alpha$ model is given by
\begin{equation}
  H = \sum_{i=1}^{3}T_{i} - T_{\mathrm{cm}} 
    + \sum_{i>j=1} \left(V^{2\alpha}_{ij} + V^{\mathrm{Coul}}_{ij}\right)
    + V^{3\alpha},
\end{equation}
where $T_i$ is the kinetic energy of $i$th $\alpha$ particle and 
the kinetic energy of the center-of-mass 
motion $T_{\mathrm{cm}}$ is properly subtracted.
The $2\alpha$ ($V^{2\alpha}$) and Coulomb ($V^{\rm Coul}$)
interactions are taken from those used in the Ref.~\cite{Fukatsu92}.
The oscillator parameter for the $\alpha$ particle
is set to be $\nu=0.26$~fm$^{-2}$, which is used to define
 the TCG basis and the forbidden states between the $\alpha$ particles.
The $3\alpha$ interaction $V^{3\alpha}$
is taken from Ref.~\cite{Ohtsubo13}
to adjust the $3\alpha$ threshold.
The mass parameter of the $\alpha$ particle and 
the elementary charge are $\hbar^2/m_{\alpha}=10.654$ MeV~fm$^2$ 
and $e^2=1.440$ MeV~fm, respectively

The wave function of the $3\alpha$ system
is described by a superposition of $K$ basis functions:
\begin{align}
  \Psi(\bm{x}) = \sum_{i=1}^{K} C_{i} \psi(A_{i},\bm{x}).
\end{align}
A set of linear coefficients $C_i$ is determined by 
solving the generalized eigenvalue problem
\begin{equation}
  \sum_{j}H_{ij} C_{j} = E \sum_{j} B_{ij} C_{j},
\end{equation}
where the matrix elements $H_{ij}$ and $B_{ij}$ are respectively defined by
\begin{align}
  H_{ij} &= \braket{\psi(A_i,\bm{x})|H|\psi(A_j,\bm{x})} \\
  B_{ij} &= \braket{\psi(A_i,\bm{x})|\psi(A_j,\bm{x})}.
\end{align}

Here two basis functions are compared:
The TCG of Eq.~(\ref{TCG.eq}) 
\begin{align}
  \psi(A_i,\bm{x})=\bar{G}(A_i,\bm{x})
\end{align}
and a symmetrized correlated Gaussian (SCG)
\begin{align}
  \psi(A_i,\bm{x})=\mathcal{S}G(A_i,\bm{x}).
\end{align}
where the $\mathcal{S}$ is the symmetrizer
to ensure the bosonic property of $\alpha$ particles,
consisting of $N!$ permutation terms that make
the CG fully symmetric under all exchanges of $\alpha$ particles.
Note that no explicit symmetrization is needed for the TCG
as it already includes the necessary symmetry in the basis function.
We again note that $\bar{G}(A,\bm{x})$
is described by the superposition of $G(A,\bm{x})$,
and thus the formulas of various matrix elements can be used
as given in Refs.~\cite{SVMbook,Suzuki08}.
As we see later, the TCG is almost free from the PFS components,
while the SCG includes a certain amount
of the PFS components, which should be eliminated properly.
To obtain physical solutions with the SCG basis functions,
the pseudopotential
\begin{equation}
  \label{eq:Pfs}
   \lambda \sum_{i>j}V_{ij}^{P}=\lambda \sum_{nlm \in \mathrm{PFS}} \ket{\phi_{nlm}(ij)} \bra{\phi_{nlm}(ij)},
\end{equation}
is added to the Hamiltonian~\cite{Kukulin78}.
Here the forbidden state wave functions
$\phi_{nlm}$ are the harmonic-oscillator wave functions
with $0S$, $1S$, and $0D$ states, i.e., $(n,l)=(0,0), (1,0)$, and (0,2),
with oscillator parameter $\nu$.
Taking the strength of the pseudopotential, $\lambda$,
large, the forbidden states can be eliminated variationally
through a superposition of many SCG states.

\section{Results}
\label{sec:resutls}

\subsection{$2\alpha$ system}
 
\begin{figure}
    \centering
    \includegraphics[width=\linewidth]{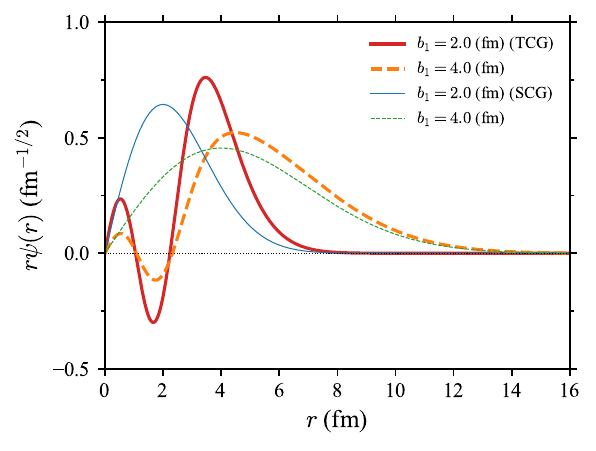}
    \caption{\label{fig:2alpha}
      Normalized TCG and SCG basis functions for $2\alpha$ system
      as a function of the distance between the two $\alpha$ particles $r$.}
\end{figure}

We first show the behavior of the TCG basis function for a $2\alpha$ system.
Figure~\ref{fig:2alpha} compares TCG and SCG basis functions
for $2\alpha$ system with relative $S$ wave.
Two parameters $\nu A_{11}=1/b_1^2$ with $b_1=2.0$ and $4.0$ fm are shown and
they are taken commonly to the TCG and SCG basis functions.
As the physical state should be orthogonal
to the forbidden $0S$ and $1S$ states,
the peak of the TCG basis functions are shifted
to far distances and two nodes appear in the internal regions.
The components of the PFS $\left<\sum_{i>j}V_{ij}^{P}\right>$
are reasonably small as 0.031 (0.006) for $b_1=2.0$ ($4.0$) fm,
while the value is 0.76 (0.21) for the SCG basis function,
indicating a significant amount of the PFS.

\subsection{$3\alpha$ system}

\subsubsection{Energy convergence}

\begin{figure}
  \centering
  \includegraphics[bb=0.000000 0.000000 270.162500 215.157500, width=\linewidth]{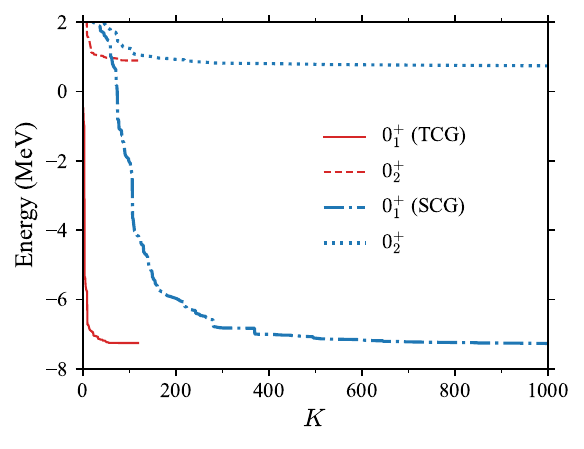}
  \caption{\label{fig:conv} Energies
    of the ground $(0_1^+)$ and first excited $0^+$ ($0_2^+$) states
    of the $3\alpha$ system
    as a function of the number of basis functions $K$
    calculated by the TCG and SCG approaches.
    The energy is measured from the $3\alpha$ threshold.}
\end{figure}

We present the numerical efficiency of
the present basis function  in this section.
Here we employ the stochastic variational method (SVM)~\cite{SVMbook}
to determine a set of the variational parameter $\left\{A_i\right\}$.
First, we generate several candidates with different variational
parameter. Practically, we take
$A_i$ in physically important regions
with the condition ${\rm det}(\Gamma-A)>0$. Expressing
$\nu A=\mathcal{R}(\theta) D$, 
where $\mathcal{R}(\theta)$ is the two-dimensional rotation matrix
with the angle $\theta$ and $D$ is a $2\times 2$
diagonal matrix given by
\begin{equation}
  D= 
  \begin{pmatrix}
    1/b_1^2 & 0 \\
    0 & 1/b_2^2
  \end{pmatrix}.
\end{equation}
In the present calculations, $b_1$, $b_2$, $\theta$ are generated
randomly in the range of $b_1,b_2<12$ fm,
and $0<\theta<2\pi$.
Candidates that has large overlap $(>0.95)$ with 
other basis functions are discarded to ensure
the linear independence of the selected basis functions numerically.
We include the candidate that gives the lowest energy among them
as the basis function and increase the total number of basis functions $K$
until the energy is converged.

Now we compare the efficiency of the TCG
and SCG plus pseudopotential approaches.
Figure~\ref{fig:conv} plots the calculated energies from the 3$\alpha$
threshold as a function of the number of basis functions, $K$.
For a fair comparison, 20 candidates at each step are generated
for both the TCG and SCG approaches.
The prefactor of the pseudopotential for the SCG approach
is taken as $\lambda=10^3$.
We see that the energy convergence with TCG is very fast
only about 100 basis functions are needed,
while SCG requires about 1000 basis states.
The energy of SCG is quite large at small $K\lesssim 100$ 
because the CG has a large overlap with the PFS,
and thus the energy becomes high due to the prefactor
of the pseudopotential.
In contrast, no such ill behavior is found for TCG because the components
of the PFS for each basis is reasonably small
about $10^{-5}$--$10^{-3}$.
As the TCG basis function already has a nodal behavior
in the internal region coming from the orthogonality to the PFS,
the basis functions at small $K$ can be used
to improve the asymptotic behavior of
the wave function, while SCG has to describe both
the internal and asymptotic behavior of the relative wave function,
leading to a large $K$ to get the converged result.  

The results are summarized in Tab.~\ref{energy.tab}.
The TCG approach gives the virtually the same energy to
that of the SCG plus pseudopotential approach
with one order of magnitude smaller basis dimension,
showing the effectiveness of the TCG approach.
The $3\alpha$ wave function with TCG contains
the component of the PFS on the order of $10^{-3}$,
which is acceptable and sufficiently small for practical use.

In Fig.~\ref{fig:conv},
we also plot the energy convergence for the $0_2^+$ state, although the state is not a bound state
but is treated as a quasi-bound state.
  As shown for the $0^+_1$ case,
  a faster convergence in the TCG approach is attained as well.
In the SCG approach, the convergence is slower than that of the ground state.
Because the $0_2^+$ state should be orthogonal to the
ground state, its full convergence will be carried out
after the PFS components in the ground state become small.
  Despite that we generate the basis states optimized for the $0^+_1$ state,
we get the energies of the $0_2^+$ state at 0.96 and 0.74 MeV
for the TCG and SCG approaches, respectively.
The components of the PFS for the $0_2^+$ state are the same order of
magnitude those obtained for the $0^+_1$ state.

Though we get much faster energy convergence in the TCG approach,
  the evaluation of the matrix element of the TCG basis is more expensive
  than the SCG basis as the number of the summation on $p$ is large,
  which is 20 times larger than the SCG basis for the $3\alpha$ system.
  However, the TCG approach still has some advantages:
  It does not require the large prefactor
  used in the pseudopotential which leads to the loss of significant digits
  in the numerical computation. While superposing non-orthogonalized basis, we face the difficulty to keep the linear independence of each basis function when the number of basis states becomes large. The TCG approach is numerically
  more stable than the SCG approach as the energy convergence
  can be achieved in the smaller basis size.

\begin{table}
 \centering
 \begin{tabular}{cccc}
   \hline\hline
        & $K$ & $E$ (MeV) & $\left<\sum_{i>j}V_{ij}^{P}\right>$\\
\hline
   TCG &120&$-7.259$ & 0.0025\\
    SCG &1000&$-7.269$ & $\approx 10^{-7}$\\
    \hline\hline
 \end{tabular}
 \caption{Number of bases, energy measured from the $3\alpha$ threshold, and the component of the PFS
   for the $0_1^+$ state of $^{12}$C.}
 \label{energy.tab}
\end{table}

\subsection{Structure of the Hoyle state}

\begin{figure*}
  \centering
  \includegraphics[width=\linewidth]{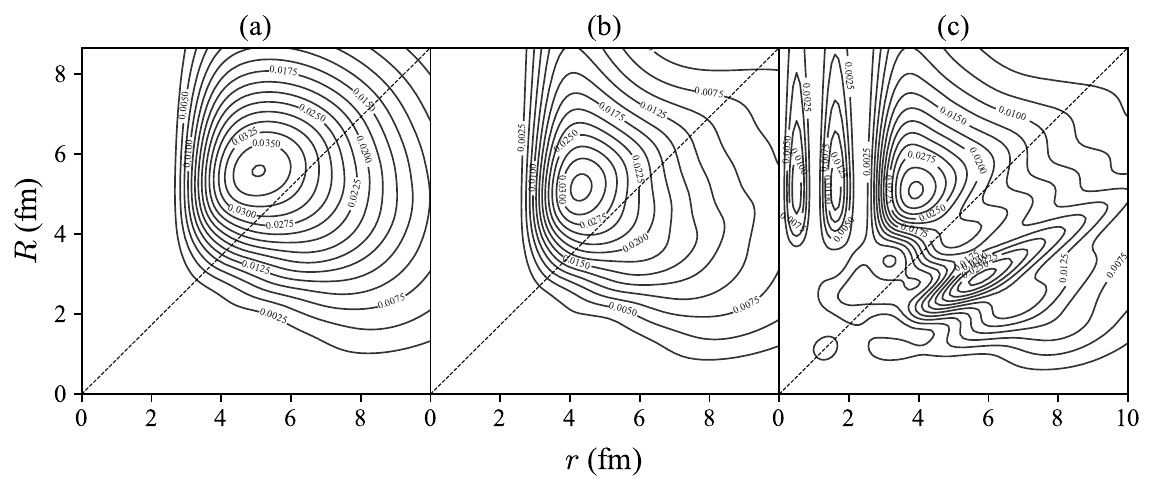}
  \caption{\label{fig:rev-tbd}
    Two-body densities of the TCG basis functions that
    give the maximum overlap with the $0_2^+$ state of $^{12}$C
    with (a) the equilateral and (b) isosceles triangle configurations,
    and (c) of the full configurations obtained by the SVM.
    The contour intervals are set to 0.0025 fm$^{-2}$.
  and }
\end{figure*}

It is interesting to investigate the structure of the Hoyle state
using TCG. It is known that the Hoyle state can be
well described by a single microscopic
$3\alpha$ cluster wave function~\cite{Tohsaki01,Funaki03}.
Following that idea, we calculate the overlap between
a TCG basis and the calculated $3\alpha$ wave function as
\begin{align}
  O(b_1,b_2)=\left.\left|\left<\bar{G}^\prime(D,\bm{x})\right|\Psi(0_2^+)\right>\right|,
\end{align}
where $\bar{G}^\prime=\bar{G}/\sqrt{\left<\bar{G}|\bar{G}\right>}$.
First, we search for the maximum value
by taking $b_2=\sqrt{3}b_1/2$, which corresponds to
geometry in that the $3\alpha$ particles are located
as an equilateral triangle. The value is $0.83$
at $(b_1,b_2)=(5.32,4.61)$ fm.
The largest overlap with the Hoyle state is found
to be 0.90 at $(b_1,b_2)=(2.30, 7.69)$ fm,
indicating an acute isosceles triangle shape,
which is consistent with the previous works~\cite{Nguyen13,Ishikawa14,Moriya21,Moriya23}.
We remark that  the maximum overlap was found to be 0.99
when the microscopic 3$\alpha$ wave function is used~\cite{Tohsaki01}.
This reduction of the overlap value is due to the approximate
orthogonality condition of Eq.~(\ref{OC2.eq}),
which can be understood by reminding 
that the spectroscopic factor of the Hoyle state
is 0.918 in the microscopic $3\alpha$ wave function~~\cite{Matsumura04}.

It is interesting to see what the optimized TCG basis looks like.
To visualize it, we calculate the two-body density (TBD), which is defined as
\begin{equation}
  \rho(r,R) = \braket{\Psi|\delta(|\bm{x}_1|-r)\delta(|\bm{x}_2|-R)|\Psi},
\end{equation}
where $r=|\bm{x}_1|=|\bm{R}_1-\bm{R}_2|$ and $R=|\bm{x}_2|=|(\bm{R}_1+\bm{R}_2)/2-\bm{R}_3|$.
Note that the TBD satisfies the normalization
\begin{align}
\int_0^\infty dr \int_0^\infty dR\,\rho(r,R)=1.
\end{align}
  Figure~\ref{fig:rev-tbd} draws
the TBDs of the TCG basis functions
with the equilateral triangle and isosceles triangle configurations
that give the maximum overlap with the Hoyle state.
The dashed line represents $R=\sqrt{3}r/2$ 
representing a ratio of $r$ and $R$ with an equilateral triangle.
Both configurations show a peak at the isosceles triangle shape,
which corresponds the dominant peak in the full calculation drawn
in Fig.~\ref{fig:rev-tbd} (c).
By taking the isosceles triangle configurations explicitly,
where the overlap with the Hoyle state is maximum,
the peak of the two-body density becomes closer
to the isosceles triangle peak in the full distribution.

Interestingly, the TCG basis with the equilateral triangle configuration
already show the peak at the isosceles triangle region.
The Pauli principle among $\alpha$ particles
generate isosceles triangle configurations,
which does not depend on the Hamiltonian employed.
The orthogonality condition on the PFS between
the $\alpha$ particles, $0S, 1S$, and $0D$ orbits,
results in strong angular momentum dependence in the relative motion and
thus the peak position of the TBD is changed to the isosceles triangle shape.
We note that a shallow potential model that does not include
the PFS~\cite{AB,Baye87,Ishikawa13,Arai18,Moriya21}
simulates this interaction-independent correlation by
strong angular momentum dependence of the intercluster potential.
In that case, the peak positions are strongly model dependent
as was shown in Ref.~\cite{Moriya21}.

\section{Conclusion}
\label{sec:summary}

To efficiently solve many-body Schr\"odinger equations involving
composite particles, we propose to use a basis function
for the orthogonality condition model
based on the microscopic cluster model approach.
We make use of the correlated Gaussian (CG) function,
which is analytically transformed to the relative wave function
of a microscopic multi-cluster system~\cite{Matsumura04}.
This transformed correlated Gaussian (TCG) function,
which is almost free from the Pauli-forbidden states,
has been applied to a few $\alpha$ systems for the first time.

Taking an example for a $3\alpha$ system, we have 
presented the efficiency of the present basis function
by comparing it with the conventional pseudopotential method
using the projection operator.
Since the basis is expressed again in a superposition of the ordinary
CG functions, its implementation is straightforward.
We show that the component of the Pauli-forbidden states is
sufficiently small in the present basis function,
allowing one to obtain stable numerical solutions
of the many-body Schr\"odinger equation with the orthogonality
conditions. The number of basis functions to get the energy convergence
becomes significantly smaller compared
to that with the conventional pseudopotential method.
Using the TCG basis,  we show that
the geometrical shape of the first excited $0^+$ state
is determined only by the Pauli principle which does not depend on
the interaction between the clusters.

In this paper,
the feasibility of the present approach has been tested
for the $3\alpha$ system.
An extension to more-$\alpha$ systems such as $^{16}$O
and $^{20}$Ne is interesting.
Also, the present study opens applications to other cluster systems involving
heavier cluster, e.g., $^{21}$Ne as $^{16}{\rm O}+\alpha+n$
and $^{24}$Mg as $^{16}{\rm O}+\alpha+\alpha$.
These applications are straightforward
as the $^{16}{\rm O}+\alpha$ potential is available~\cite{Arai18}.

\section*{Acknowledgments}

This work was in part supported by JSPS KAKENHI Grants
Nos.\ 18K03635 and 22H01214, and
the National Natural Science Foundation of China under Grant No. 12175042.
We acknowledge the Collaborative Research Program 2022, 
Information Initiative Center, Hokkaido University.

\end{document}